\documentclass[aip,jcp,reprint]{revtex4-1}

\usepackage[utf8]{inputenc}
\usepackage[fleqn]{amsmath}
\usepackage{amsfonts}
\usepackage{amssymb}
\usepackage{amsthm}
\usepackage{amsopn}
\usepackage{upgreek}
\usepackage{bm}
\usepackage{epsfig}
\usepackage{relsize}

\newcommand{\Zid}{Z_{\text{id}}}
\newcommand{\Zin}{Z_{\text{in}}}
\newcommand{\qc}{q_{\text{c}}}
\newcommand{\qd}{q_{\text{d}}}
\newcommand{\be}{\beta\epsilon}
\newcommand{\bbe}{\beta\bar\epsilon}
\newcommand{\bte}{\beta\tilde\epsilon}
\newcommand{\bw}{\beta w}
\newcommand{\behalf}{\beta\epsilon_{50}}
\newcommand{\barzEV}{\bar{z}_{E,V}}
\newcommand{\barAEV}{\bar{A}_{E,V}}
\newcommand{\barBEV}{\bar{B}_{E,V}}
\newcommand{\bteEV}{\bte_{E,V}}
\newcommand{\av}{\{\epsilon_b\}}

\newcommand{\kB}{k_{\text{B}}}
\newcommand{\Vmax}{V_{\text{max}}}

\renewcommand{\eqref}[1]{Eq.~(\ref{#1})}

\newcommand{\figref}[1]{Figure~\ref{#1}}
\newcommand{\figsref}[1]{Figures~\ref{#1}}

\newcommand{\refcite}[1]{Ref.~\onlinecite{#1}}

\begin{document}

\newcommand{\addresscambridge}{Department of Chemistry, University of Cambridge, Lensfield Road, Cambridge CB2 1EW, United Kingdom}

\title{Theoretical prediction of free-energy landscapes for complex self-assembly}
\author{William M.~Jacobs}
\affiliation{\addresscambridge}
\author{Aleks~Reinhardt}
\affiliation{\addresscambridge}
\author{Daan~Frenkel}
\affiliation{\addresscambridge}
\date{\today}

\begin{abstract}
We present a technique for calculating free-energy profiles for the nucleation of multicomponent structures that contain as many species as building blocks.  We find that a key factor is the topology of the graph describing the connectivity of the target assembly.  By considering the designed interactions separately from weaker, incidental interactions, our approach yields predictions for the equilibrium yield and nucleation barriers.  These predictions are in good agreement with corresponding Monte Carlo simulations.  We show that a few fundamental properties of the connectivity graph determine the most prominent features of the assembly thermodynamics.  Surprisingly, we find that polydispersity in the strengths of the designed interactions stabilizes intermediate structures and can be used to sculpt the free-energy landscape for self-assembly.  Finally, we demonstrate that weak incidental interactions can preclude assembly at equilibrium due to the combinatorial possibilities for incorrect association.
\end{abstract}

\maketitle

Building nanostructures out of multiple, distinct components offers enormous possibilities for high-fidelity manufacturing at the molecular level. Such `addressable' structures are fundamentally different from conventional crystals or ordered clusters, since every building block is distinct and thus occupies a specific location in the target structure.  Because the interactions between building blocks are specified independently, it is possible to design finite-sized, three-dimensional structures that assemble nearly error-free.\cite{hormoz2011design,halverson2013dna,hedges2014growth,zeravcic2014size}  An experimental proof-of-principle can be found in self-assembling DNA `tiles,' which use the hybridization of complementary DNA sequences to construct complex structures consisting of hundreds of subunits from a single soup of monomers.\cite{ke2012three}  Simulation results have shown that such one-pot self-assembly can succeed with highly simplified model subunits that lack the molecular details of DNA tiles, suggesting that similar design strategies should be widely applicable.\cite{reinhardt2014numerical}

In this communication, we present a simple and efficient method for predicting the free-energy landscape for the self-assembly of addressable structures directly from the graph describing the target structure.  Our approach allows us to predict the assembly yield and nucleation barriers quantitatively for any multicomponent structure in which directional, designed interactions stabilize the target assembly.  We demonstrate the accuracy of this method by comparing our predictions with lattice Monte Carlo (MC) simulations of a DNA-tile nanostructure.\cite{reinhardt2014numerical} The good agreement suggests that our approach can be directly applied to any experimental system with designable interactions.

\begin{figure}[t!]
  \includegraphics{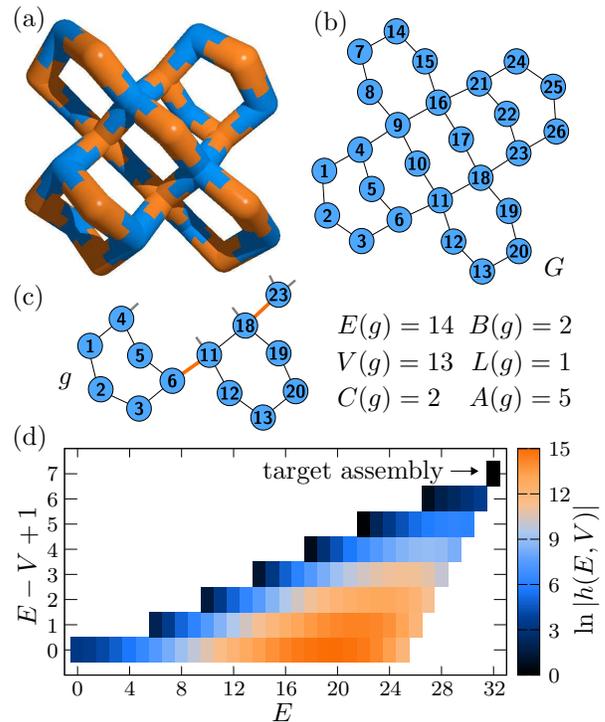}
  \caption{(a) An example three-dimensional DNA-tile structure in which all 26 subunits are distinct.  For comparison with simulations,\cite{reinhardt2014numerical} the monomers are constrained to a cubic lattice with the pegs oriented toward one of the four nearest-neighbor sites.  Every designed interaction between adjacent subunits is distinct, although the subunits can associate in any of the three possible dihedral configurations.  (b) The connectivity graph, $G$, representing the designed interactions between arbitrarily labeled subunits.  (c)  A subgraph, $g$, of the target structure and its topological properties, as defined in the text.  Bridges are highlighted in orange, and edges adjacent to $g$ are shown in gray.  (d) The logarithm of the `density of states' of fragments with $E$ edges and $V$ vertices.  The vertical axis indicates the number of linearly independent cycles in fragments within each set.}
  \label{fig:structure}
\end{figure}

Our method relies on the observation that the designed interactions in the target structure are typically much stronger than any incidental associations between subunits that should not be connected in the final assembly.  These designed interactions can be represented by a graph, $G$.  The vertices of this graph are all unique, since each vertex corresponds to a particular subunit.  The edges indicate designed interactions that stabilize the target assembly.  An example 26-subunit structure and its associated connectivity graph are shown in \figref{fig:structure}.  Initially, we consider only designed interactions that are correct for error-free assembly of the target structure.  Later, we shall discuss how to account for incidental interactions.

In order to describe the assembly of the target structure, we must consider the relative stability of every possible correctly bonded partial structure (`fragment').  In terms of our graph, a fragment corresponds to a connected subgraph of~$G$.  Fragments that differ only in the labels of their vertices are distinct because the vertices of $G$ are all unique.  Even though the total number of fragments grows exponentially with the number of edges in the target structure, limiting our attention to correctly bonded clusters ensures that the set of fragments is finite.

We can deal with the enormous number of partial structures by grouping the fragments, $\{g\}$, into sets with the same number of edges, $E(g)$, and vertices, $V(g)$.  While all fragments in a set $h(E,V)$ may not share the same topology, they are likely to have similar thermodynamic properties.  We then count the number of fragments in each set, $|h(E,V)|$, statistically by applying the Wang--Landau flat-histogram algorithm\cite{wang2001efficient} to the state space of connected subgraphs of $G$.  In this algorithm, a stochastic trajectory hops among graphs in the fragment state space, making transitions between fragments that differ by a single edge.  For every visited fragment $g$, we calculate the number of `bridges' (edges that, if cut, would break $g$ into two disconnected subgraphs), $B(g)$, and `leaves' (bridges that attach a single vertex to the rest of $g$), $L(g)$ (\figref{fig:structure}c).  The removal of a bridge that is not also a leaf would disconnect $g$ into two nontrivial graphs and is thus not allowed.  We also find all edges that are adjacent to $g$ in $G$ and denote the number of such edges by $A(g)$.  In order to obey detailed balance in this state space, randomly proposed edge additions or removals to transition between graphs $g$ and $g'$ are accepted with probability
\begin{equation}
  \label{eq:acceptance_probabilities}
  p\!\!_{\substack{\text{add}\\\text{remove}}}\!\!(g \rightarrow g') = \min \left[1, \; \frac{|h(E(g),V(g))|}{|h(E(g'),V(g'))|} \frac{n_{\scriptscriptstyle \pm}(g)}{n_{\scriptscriptstyle \mp}(g')} \right],
\end{equation}
where ${n_{\scriptscriptstyle +}(g) \equiv A(g)}$ and ${n_{\scriptscriptstyle -}(g) \equiv E(g) - B(g) + L(g)}$.  When leaves are added or removed, the accompanying free vertex is attached or discarded from $g'$ as well.  By updating $|h(E,V)|$ according to the algorithm described in \refcite{wang2001efficient}, we can calculate the number of fragments in each set efficiently and to arbitrary precision.

The resulting `density of states' is an intrinsic property of the target structure that determines the most important features of its assembly at equilibrium.  For reasons that will soon become clear, we have organized the $\{E,V\}$-sets in \figref{fig:structure}d according to the number of edges and the number of linearly independent cycles, ${C \equiv E - V + 1}$.  Conveniently, $|h(E,V)|$ need only be calculated once for a particular target structure.

In a dilute solution with many copies of each component, we can treat the mixture as an ideal solution of fragments.  It is expedient to work in the grand-canonical ensemble, in which monomers of each component can be exchanged with an infinite reservoir at constant chemical potential.  For simplicity, we assume that these chemical potentials are chosen such that each type of monomer is present in the same concentration, $\rho$, but in general, these concentrations can be different. Because all fragments $\{g\}$ are in chemical equilibrium with the monomer reservoirs, the dimensionless grand potential is simply the sum of the fugacities $\{z_g\}$ of all fragments $\{g\}$: ${-\ln \Xi = -\sum_g z_g}$ (see, e.g.~\refcite{hansen2006theory}).  The sum over fragments can then be replaced by a sum over sets of fragments $h(E,V)$, where each term is accompanied by the average fugacity of the fragments in the set, $\barzEV$:
\begin{equation}
  \label{eq:lnXi}
  \Zid \equiv \ln \Xi = \sum_{E,V} | h(E,V) | \,\barzEV.
\end{equation}

For every fragment, the associated fugacity can be determined directly from the fragment graph.  First, there is an attractive contribution due to the designed binding energies: ${-\beta\sum_{b \in \mathcal{E}(g)} \epsilon_b}$, where ${\beta \equiv 1/\kB T}$, $\mathcal{E}(g)$ is the edge set of $g$ and $\{ \epsilon_b \}$ are the bond energies.  Second, there is an entropic cost to bring monomers into contact, ${V\ln\rho}$.  Finally, there is an entropic penalty due to the loss of rotational entropy of bonded subunits.  Ignoring excluded volume interactions, the formation of a bond forces the interaction sites on the monomers to face one another, reducing the dimensionless entropy by $\ln\qc$, where $\qc$ is the coordination number of the lattice.  Each bond that does not correspond to a bridge costs a further $\ln\qd$ of entropy, where $\qd$ is the number of dihedral angles possible for two bonded monomers on the lattice.  The average fugacity of fragments in the set $h(E,V)$ is thus
\begin{eqnarray}
  \label{eq:lnzEV}
  \ln\barzEV &=& E \bteEV + V \ln \rho \\
  &\quad& - (V - 1) \ln \qc - (V - \barBEV - 1) \ln \qd, \nonumber
\end{eqnarray}
where $\bte$ is the absolute value of the mean binding energy.  The dimensionless dihedral entropy ${\barBEV \ln\qd \equiv \ln \left\langle \qd^{B(g)} \right\rangle_{g \in h(E,V)}}$ is also an intrinsic property of $G$ and is easily computed with a stochastic calculation in the fragment state space.\footnote{To calculate $\barBEV$, $\barAEV$ and $\bte_{E,V}$, we perform a biased Monte Carlo calculation using the pre-calculated density of states $|h(E,V)|$ and the acceptance probabilities given in \eqref{eq:acceptance_probabilities}.}

\begin{figure*}
  \includegraphics{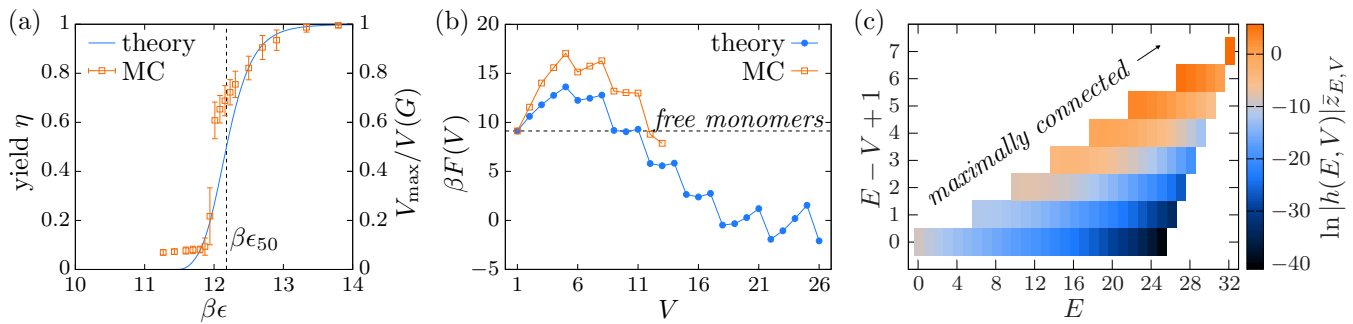}
  \caption{(a) Comparison of the predicted equilibrium yield of the example structure and the largest stable cluster size, $\Vmax$, obtained from MC simulations with a single copy of each subunit.  All monomers are present in the same concentration, ${\rho = 62^{-3}}$.  (b) Comparison of the predicted and simulated free-energy profiles with ${\be = 12.07}$.  (c) The statistical weights in \eqref{eq:lnXi} at~$\behalf$.  Details of the MC simulation methods used in (a) and (b) are provided in \refcite{reinhardt2014numerical}.}
  \label{fig:designed_only}
\end{figure*}

We first consider the case where all designed bond energies have equal magnitude, i.e.~${\bteEV = \be\, \forall E,V}$.  We define the yield, $\eta$, to be the grand-canonical average of the fraction of all clusters in solution that match the target structure,
\begin{equation}
  \eta \equiv \frac{\langle N_G \rangle}{\sum_g \langle N_g \rangle} = \frac{z_G}{\Zid},
\end{equation}
where $N_g$ is the number of copies of fragment $g$.  Our prediction for the equilibrium yield of the structure described in \figref{fig:structure} is shown in \figref{fig:designed_only}a.  Quite strikingly, the transition from zero to nearly 100\% yield over approximately $2~\kB T$ suggests highly cooperative assembly.  Adjusting the monomer concentration simply shifts the yield curve, with the 50\%-yield bond strength, $\behalf$, changing linearly with $\ln\rho$.  Increasing the bond strength beyond ${\be \simeq 14}$ results in perfect assembly because only designed interactions are considered at this point.  The predicted yield curve coincides remarkably well with the largest stable cluster observed in MC simulations.

The correspondence between the intrinsic properties of the connectivity graph and the equilibrium self-assembly of the target structure is immediately apparent from the free-energy profile shown in \figref{fig:designed_only}b.  Because we are interested in the progress toward complete assembly starting from any subset of components, the relevant free energy is a sum over all fragments consisting of $V$ monomers,
\begin{equation}
  \beta F(V) \equiv -\ln \sum_E | h(E,V) | \,\barzEV.
\end{equation}
Exactly seven free-energy barriers, each corresponding to a linearly independent cycle in $G$, must be crossed in order to assemble the example structure from free monomers in solution.  Again, we find good agreement with the results of MC simulations.\footnote{In lattice simulations with a single copy of each subunit, we define ${\beta F(V) + \ln\left[\rho V(G)\right] \equiv -\ln \langle N_V / V(G) \rangle + V \mu^*(V)}$,~where $N_V$ is the number of clusters with $V$ subunits and ${\mu^*(V) \equiv \ln \{[V(G) + 1 - V]/V(G)\}}$ approximates the change in the chemical potential due to monomer depletion.}  The ascent on approach to each barrier results from the recruitment of additional subunits by single bonds.  Each steep descent corresponds to the completion of a new cycle, in which the additional bond compensates for the loss of rotational and translation entropy.  The critical nucleus is ${V = 5}$, since six subunits are required to form a cycle in the example structure.  Beyond the critical nucleus, the number of monomers required to complete additional cycles can be determined directly from $|h(E,V)|$.  Unsurprisingly, fragments with the greatest number of bonds per monomer are most stable and thus most likely to form at equilibrium, as shown in \figref{fig:designed_only}c.

\begin{figure*}
  \includegraphics{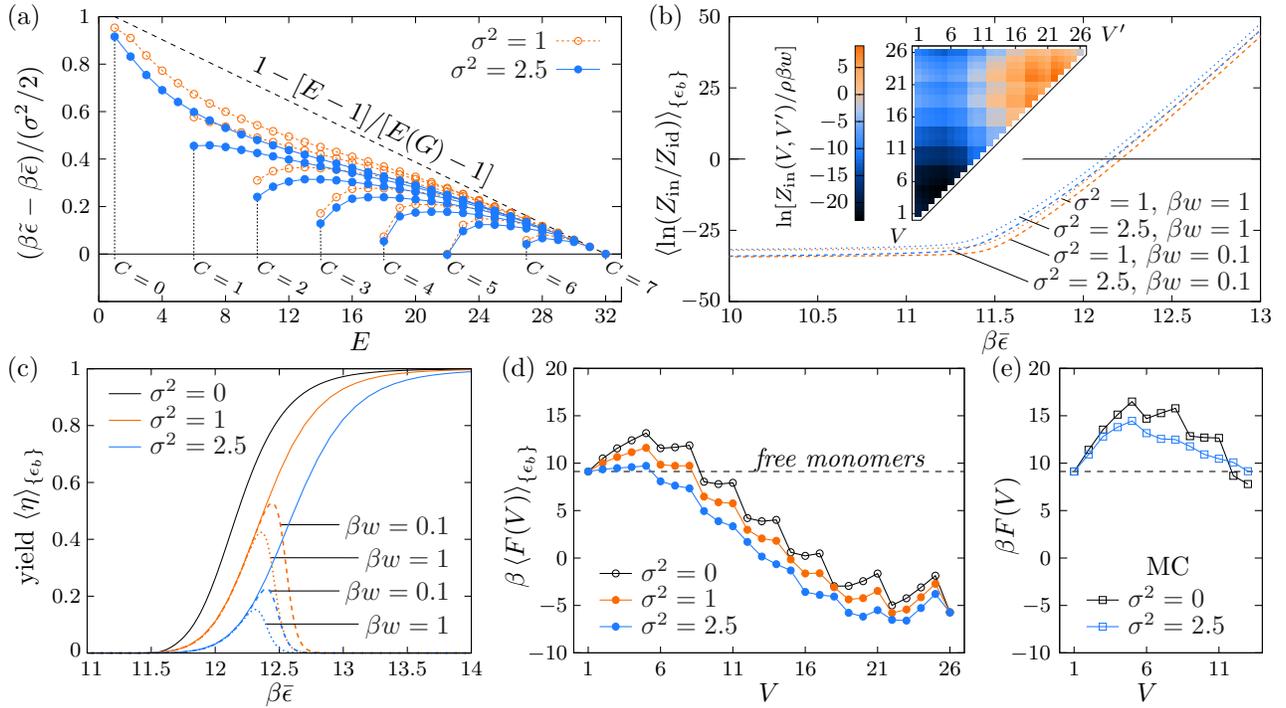}
  \caption{(a) Effective bond energies for each set of fragments, $\bteEV$, averaged over Gaussian-distributed designed interactions with variance~$\sigma^2$.  (b) The ratio of the contributions to the grand potential due to incidental interactions, $\Zin$, and designed interactions, $\Zid$.  (b,\textit{inset}) The statistical weights of incorrectly formed dimers of fragments with $V$ and $V'$ vertices at $\behalf$ for the case of identical bond energies.  (c) Predicted average yields given polydisperse energies, with and without incidental interactions.  (d) Predicted average free-energy profiles with polydisperse energies.  (e) Comparison of MC simulations with identical bond energies and a particular set of quenched energies.}
  \label{fig:polydisperse_incidental}
\end{figure*}

But what if the strengths of all designed interactions are not identical?  If the designed interactions are instead chosen randomly from a distribution with mean $-\bbe$ and a finite variance $\sigma^2$, then we should expect the thermodynamic properties of the self-assembling system to fluctuate as well.  Because the bond energies do not change during assembly, we must treat the $\{\epsilon_b\}$'s as \textit{quenched random variables} when taking thermal averages.  The mean bond energy of the target structure is clearly self-averaging, since ${\tilde\epsilon(E(G),V(G)) \rightarrow \bar\epsilon}$ by the central limit theorem.  In the case of partial structures, however, the effective mean bond energy $\bteEV$ depends on the number of fragments in the set $h(E,V)$.  Because self-averaging is only meaningful in the context of free energies that are extensive in $E(G)$,\cite{fischer1993spin} the effective mean bond energy is
\begin{equation}
  \label{eq:mean_energies}
  \bteEV \equiv \left.\mathlarger{\Biggl\langle} \frac{1}{E} \ln \left\langle \exp \! \sum_{b \in \mathcal{E}(g)} \!\! \be_b \right\rangle_{\!\!\! g \in h(E,V)} \mathlarger{\Biggr\rangle}\right._{\!\!\!\! \av} \!\!\!\!. \,\,
\end{equation}
The inner $\langle\cdot\rangle$ is a thermal average over all fragments with \textit{fixed} bond energies corresponding to each edge in $G$.  Assuming a Gaussian distribution of independently chosen $\{\beta\epsilon_b\}$'s, we would find ${\ln\langle \cdot \rangle_g / E \rightarrow \bar\epsilon + \frac{1}{2}\sigma^2 \{1 - [E - 1]/[E(G) - 1]\}}$ if all combinations of edges were allowed in the fragments.  The requirement for subgraphs to be connected, however, introduces additional correlations among the bond energies in $h(E,V)$.  The outer $\langle\cdot\rangle$ averages over the quenched random interactions, which introduces an additional correction due to the finite number of bonds.  The effective mean bond energies of the example structure, with ${\sigma^2 = 1}$ and ${\sigma^2 = 2.5}$, are shown in \figref{fig:polydisperse_incidental}a.

The physical consequence of this analysis is that polydispersity in the designed interaction energies tends to stabilize partially formed structures.  This effect originates from the exponential weighting of the bond energies, which is most significant when the fragments are small and the number of fragments in a set is large.  Because relatively few fragments with a given number of edges have many cycles, energy polydispersity also tends to stabilize `floppy' fragments.  As a result, the yield curves broaden and shift to higher $\bbe$ as the variance in the interaction energy distribution is increased (\figref{fig:polydisperse_incidental}c).  Energy polydispersity also rounds off the peaks of the free-energy barriers, diminishing the signature of the cycle-dependence for very broad interaction energy distributions (\figsref{fig:polydisperse_incidental}d-e).

Finally, we can estimate the thermodynamic consequences of incidental interactions between subunits.  In terms of the connectivity graph, incidental interactions occur between vertices in two fragments that both have adjacent edges but do not share an edge in $G$.  Assuming that the mean incidental interaction strength, $\bw$, is significantly weaker than $\bbe$, we can use a high-temperature expansion to estimate the effect of incidental interactions in a reference state of ideal fragments.\cite{hansen2006theory}  The additional possibilities for binding result in a decreased dimensionless grand potential, ${-\ln\Xi = -\Zid - \Zin - \mathcal{O}(w^2)}$.  The second term in this expansion accounts for all possible incidental interactions between fragments in the ideal reference state,\footnote{Assuming ideal polymers, binding reduces the dimensionless rotational entropy of one fragment by $\ln\qc$.  One fragment loses $\ln\qd$ of dihedral entropy for a single contact, while further contacts cost $2\ln\qd$.  Although this formula is approximate, $\Zin/\Zid$ is relatively insensitive to the precise form of the coefficients in \eqref{eq:Zin}.}
\begin{equation}
  \label{eq:Zin}
  \Zin \simeq \sum_{g,g'} z_g z_{g'} \!\!\!\!\!\! \sum_\lambda^{\min(A_g, A_{g'})} \!\!\! {A_g \choose \lambda} {A_{g'} \choose \lambda} \lambda! \frac{2^{-\delta_{gg'}} \lambda \rho \beta w}{\qc \qd^{2\lambda - 1}},
\end{equation}
where the sum over $\lambda$ accounts for the many incidental binding opportunities between the `multivalent' fragments.\cite{martinez2011designing}  $\Zin$ can be rewritten as a sum over ${\{E,V;E',V'\}}$, with ${\barAEV \equiv \langle A_g \rangle_{g \in h(E,V)}}$ determined from a stochastic calculation in the fragment state space.

For successful assembly to occur at equilibrium, incidental interactions must be less probable than designed interactions, such that ${\Zin \lesssim \Zid}$.  \figref{fig:polydisperse_incidental}b shows that ${\Zin / \Zid}$ increases sharply as soon as multimeric fragments become populated (${\bbe \simeq 11.3}$).  As a result, the yield quickly drops to zero, even with extraordinarily weak incidental interactions (\figref{fig:polydisperse_incidental}c).   Although the point at which $\Zin$ exceeds $\Zid$ is relatively insensitive to both $\bw$ and $\sigma^2$, the equilibrium assembly window narrows rapidly since larger clusters, which are numerous under conditions of high yield, are also the most susceptible to aggregation (\figref{fig:polydisperse_incidental}b,\textit{inset}).  Incidental interactions therefore present a fundamental thermodynamic constraint for successful self-assembly at equilibrium.

The method presented here is generally applicable to any addressable structure that is stabilized by specific, directional interactions.  By considering the complete set of on-pathway assembly intermediates, this approach reveals how the fundamental topological properties of the connectivity graph determine the free-energy landscape for self-assembly.
This method is also spectacularly efficient: the calculations required to generate \figref{fig:designed_only}b, for example, were six orders of magnitude faster than the corresponding MC simulations, and the fragment density of states can be reused to compute free-energy profiles at any temperature and concentration.
Futhermore, generalizations of this method may be used to study nucleation in complex crystals and the assembly pathways of heteropolymers with designed native structures.
This graph-based approach therefore points the way toward improved designs for a broad class of self-assembling nanostructures.

W.M.J.~acknowledges support from the Gates Cambridge Trust and the National Science Foundation Graduate Research Fellowship under Grant No.~DGE-1143678.  D.F.~acknowledges European Research Council Advanced Grant 227758 and Engineering and Physical Sciences Research Council Programme Grant EP/I001352/1.  Research carried out in part at the Center for Functional Nanomaterials, Brookhaven National Laboratory, which is supported by the US Department of Energy, Office of Basic Energy Sciences, under Contract No.~DE-AC02-98CH10886.

\end{document}